\documentstyle[12pt]{article}
\newcommand{\newc}{\newcommand}
\newc{\beq}    {\begin{equation}}
\newc{\eeq}    {\end{equation}}
\newc{\beqa}    {\begin{eqnarray}}
\newc{\eeqa}    {\end{eqnarray}}
\newc{\ba}    {\begin{array}}
\newc{\ea}    {\end{array}}
\newc{\st}    {\stackrel}
\newc{\f}    {\frac}

\def\NPB{{\em Nucl. Phys.} {\bf B}}
\def\PLB{{\em Phys. Lett.}  {\bf B}}
\def\PRL{\em Phys. Rev. Lett. }

\def\PRD{{\em Phys. Rev.} {\bf D} }

\def\PRP{{\it  Phys. Rep.} }

\begin{document}

\renewcommand{\baselinestretch}{2}
\large
\normalsize
\noindent

\begin{titlepage}
\title{ FERMION SCATTERING AT A PHASE WAVE }
\author{ Jae-weon Lee
 and In-gyu Koh \\ \it Department of Physics,
 \\ \it Korea Advanced Institute of Science and Technology,
\\   \it 373-1, Kusung-dong, Yusung-ku, Taejeon, Korea  \\
\\
}

\setcounter{page}{1}
\date{}
\maketitle
\vspace{-9cm}
\vspace{10cm}


\hspace{.5cm}We study fermion reflection  
at a phase wave which is formed during  a bubble collision
in a first order phase transition.
We calculate the reflection and the transmission
  coefficients by solving
the Dirac equation with the phase wave background. 
Using the results we analyze
the damping and the velocity of the wave.

\vspace{1cm}
PACS number(s): 12.15.Ji, 98.80.Cq

\vspace{4cm}
\maketitle

\end{titlepage}
\newpage

\vskip 5mm
There have  been  many studies on the cosmological roles of the
 first-order phase transition which proceeds by nucleation and
collisions of vacuum bubbles\cite{bubble}, especially
in some inflation models\cite{inflation}.  

 If a global or a local symmetry is broken at the transition,
there could be a phase wave, which has not been  studied fully.
In this paper, we study a fermion scattering at the phase wave.

At the nucleation, due to a finite correlation length,
the two true vacuum bubbles that are separated by more than
the correlation length or the horizon distance may have different phases.
Such a possibility is a key ingredient for the Kibble mechanism\cite{kibble}.

If there is a phase difference between the two colliding bubbles,
 a pair of phase waves occur and  propagate into each bubble.
 The phase wave \cite{hawking,hindmarsh} is a mechanism by which
the bubbles get their new phase values determined by
the so-called ``geodesic rule'' in a first 
 \cite{hindmarsh} or a second\cite{second} order phase transition.

The energy of the colliding bubble wall turns into that 
of the modulus wall and the phase wave.
If the phase difference of the two bubbles is of order 1, the
phase wave carries away most of the energy of the bubble wall
 \cite{hawking}.
So the phase wave can be a very energetic phenomenon
 as the bubble wall is.

Furthermore, investigating the particle scattering at a moving 
bubble wall is important to study the bubble kinetics
and cosmology.
For example, to calculate the velocity and the width 
of the electro-weak bubbles\cite{velocity} and  
the CP violating charge transport rate  by the wall 
for baryogenesis
\cite{kaplan},
one should know the reflection coefficient of fermions({\it e.g.} top quarks)
scattering at the wall.

However, there have been a few works about the dynamics and cosmological
roles of the phase wave compared with the bubble walls.

So, in this paper, we investigate  the interaction between
the phase wave and fermions.
In the cosmological phase transition, a propagating phase wave may
collide with the fermions in the plasma.

Consider a complex scalar field  $\phi(x)=\rho(x) exp(i\theta(x))$ 
 whose Lagrangian density is given by
\beq
{\cal L}=\frac{1}{2}\partial_{\mu} \phi^* \partial^{\mu} \phi-V(\rho),
\eeq
where $V(\rho)$ is a potential appropriate for
 a first-order phase transition.

Since the potential  depends only on the amplitude $\rho$,
 one may choose an arbitrary phase for the minimum of the potential,
 $\langle\phi\rangle$.
 Of course, this is
due to the global $U(1)$ symmetry the theory possesses.
In general, we can make the true vacuum to lie along the direction
of the real part of $\phi$, because the phase of $\phi$ can be
absorbed by redefining the phase of the fermion field. 
However, if $\langle \phi \rangle$ is space-time
dependent, we can not globally rotate out the phase
which may play physical roles.

From ${\cal L}$ we obtain the equation of motion for $\theta$;
\beq
\partial^\mu \partial_\mu \theta+\frac{2}{\rho}\partial^\mu\theta
 \partial_\mu \rho =0.
\label{theta}
\eeq

Without loss of generality, we can choose the two phases
for the colliding bubbles as zero and $2\triangle\theta(<\pi)$,
respectively . 
Then, by the geodesic rule, the overlapping region of the two bubbles
 should have
$\triangle\theta(<\pi/2)$.

We consider the case where the width
of the phase wave ($d$)  is much
 smaller than the radius of curvature ($r$) of the phase wave.
In this case we can treat our problem
as one dimensional.
We choose $z$ axis  as the direction normal to the
plane of the phase wave, and assume  
that momentum of the incident fermion is parallel to 
the $z$ axis. 
The solution for the more general momentum can be found by 
appropriately Lorentz boosting the one dimensional solution.

A possible solution of eq.(\ref{theta})  is
\beq
\theta(z,t)=\triangle\theta[1+ \frac{1}{2}f(z+vt) 
+\frac{1}{2}f(z-vt)],
\eeq
where $f(z)$ is  some profile function which has 
an asymptotic value $\pm1$ at $z=\pm\infty$, respectively.

The second term indicates the left moving phase wave($-z$ direction)
with velocity $v$ relative to the plasma,
 while the third term is the right moving one ($+z$ direction).
Since $\partial_\mu\rho=0$ for the phase wave, $v$ is the speed of
light $c$.
However, eq.(\ref{theta}) ignored the pressure by the
fermion scattering which is just what we are going to
study in this paper.
So we assume $|v|<c$.

To calculate the reflection coefficient of the fermion
 we choose the rest frame of the left moving phase wave ($v=0$)
 and assume that the incident fermion
comes from the left. (See Fig.1.) 
When the typical wave length and the mean free path
 of the fermion are much larger than
the phase wave width $\sim 1/m_0$,   we can    
 approximate $f(z)$ as $2\Theta(z)-1$ with
step function $\Theta(z)$,i.e.,
 $\theta(z)=\triangle \theta $ when $z \geq 0$ and
zero for $z<0$.   

Consider a Lagrangian density for the fermion $\Psi$
\beq
{ L_\Psi}= \bar{\Psi}_L i\slash{\hskip-0.25cm \partial}
\Psi_L +\bar{\Psi}_R i
\slash{\hskip-0.25cm \partial}\Psi_R
-(h \bar{\Psi}_L\Psi_R \phi+ h^*\bar{\Psi}_R\Psi_L \phi^* )
\eeq
with the (real) Yukawa coupling constant $h$.
In the rest frame of the phase wave, $\Psi$ acquires
position dependent mass $m(z)=h <\phi(z)>$. 
Therefore we get the equation of motion for $\Psi$ from ${ L_\Psi }$;
\beq
(i\slash{\hskip-0.25cm \partial}  - m(x) P_R  - m^*(x) P_L ) \Psi=0,
\label{dirac}
\eeq
where $P_{R,L}$ are the chirality projection operators.

We adopt the following ansatz\cite{ayala,funakubo}
\beqa
\Psi(t,z)&=&(i\slash {\hskip-0.25cm \partial}+m^*(z) P_R+m(z) P_L) e^{-i\sigma Et}\psi( k,z) \nonumber \\
&=&(\sigma E \gamma^0+i\gamma^3\partial_z +m_R(z)-i m_I(z)\gamma_5 ) e^{-i\sigma Et}\psi( k,z)
\label{solution},
\eeqa

where a field $\psi$ satisfying following 
the Klein-Gordon-like equation is introduced;
\beq
(E^2+\partial_z^2-|m(z)|^2+i m'_R(z)\gamma^3-m'_I(z)
\gamma_5\gamma^3)\psi(z)=0.
\label{KG}
\eeq
Here prime denotes the derivative with $z$,
 $m(z)=m_R(z)+i m_I(z)$ and
 $\sigma$ is $+1$ and $-1$ for fermion and antifermion, respectively.

If $m'(z)\neq0$, there is an additional potential term 
proportional to the spatial change of $m(z)$, which is the
origin of the scattering.
For our case, there is a delta function type potential  at $z=0$.

Let us find momentum eigen states.
We expand $\psi$ in the eigenspinors of $\gamma^3$ with
eigenvalue $\pm i$ satisfying
$\gamma^3u^s_\pm =\pm i u^s_\pm$,
where $u^1_\pm=1/\sqrt{2}(1,0,\pm i,0)^T$ and
 $u^2_\pm=1/\sqrt{2}(0,1,0,\mp i)^T$ with spin index $s=1,2$.
(We use here the same conventions for the $\gamma$ matrices and
$u^s_\pm$ in ref.\cite{ayala}.)
Note that 
$\gamma^0u^s_\pm=u^s_\mp$. 

Since $m(z)=m_0 e^{i\theta(z)}$,
 $m_R=m_0 $ and $m_I=0$
for $z < 0$ (region $I$)
and
 $ m_R(z)=m_0 cos(\triangle\theta)\equiv m_c$ and
 $m_I(z)=m_0 sin(\triangle\theta)\equiv m_s$
 for $z\geq 0$ (region $II$). 

From now on we will consider the case of particle($\sigma=1$).
For $z\neq0$, $m'(z)=0$ and eq.(\ref{KG}) becomes
a free Klein-Gordon equation.
 Then, the most general right moving solution to eq.(\ref{KG}) can be 
represented as a combination of $u^s_+$ and
$u^s_-$,.i.e., 
\beq
 \psi=e^{ikz}\displaystyle 
\sum_s \sum_\pm C^s_\pm u^s_\pm,
\eeq
where $k=\sqrt{E^2-m_0^2}$ and $C^s_\pm$ are some constants.

Inserting $\psi$ into
 eq.(\ref{solution}) we get $\Psi$ for $z<0$; 
\beq
\Psi=\displaystyle e^{-iEt+ikz}\sum_s\{
C^s_+(Eu_-^s+(m_0-ik)u^s_+) 
+C^s_-(Eu_+^s+(m_0+ik)u^s_-)\}.
\eeq
These terms are not independent in each other in $u^s_\pm$ basis, so we can
set $C^s_-=0$ without loss of generality. 
Therefore,
\beq
\displaystyle \psi(z)=e^{ikz}\sum_s C^s_+ u^s_+.
\label{psi0}
\eeq
Similarly,  for the left moving wave we can choose
$\Psi$ proportional to $ (Eu_+^s+(m_0-ik)u^s_-)$.
So for the region $I$, the general solution can be given by
\beqa
\Psi_{I}=\displaystyle e^{-iEt}\sum_s
C^{s}_+\{e^{ikz}[(m_0-ik)u^s_+ +Eu^s_-] \nonumber \\
+R_se^{-ikz}[Eu^s_+ +(m_0-ik)u^s_-]\}.
\eeqa
The first term is  a incoming wave   and 
the second term is a reflected one.
In the region $II$ only a transmitted wave exists; 
\beq
\Psi_{II}=\displaystyle e^{-iEt+ikz}\sum_s\  
T_sC^{s}_+\{(E-m_s(-)^s)u^s_-+(m_c-ik)u^s_+)\}.
\eeq
Here, $R_s$ and $T_s$ are some spin-dependent coefficients.

Now, equating $\Psi_I$ and $\Psi_{II}$ at $z=0$ and
comparing each coefficients of $u^s_\pm$, we get two
algebraic equations for $R_s$ and $T_s$;
\beqa
R_s&=& \frac{-E+(E-(-)^sm_s)T_s}{m_0-ik},  \nonumber \\ 
T_s&=& \frac{m_0-ik+ER_s}{m_c-ik},  
\eeqa
 which has a solution
\beqa
R_s&=& \frac{1}{D}[E(m_0-m_c)-(-)^sm_s(m_0-ik)], \nonumber \\
T_s&=& \frac{1}{D}[(m_0-ik)^2-E^2],
\label{RT}
\eeqa
where $D=(m_c-ik)(m_0-ik)-E(E-(-)^sm_s)$.


From $\Psi_I$ and $\Psi_{II}$ 
we get a vector current $j^3=\bar{\Psi}\gamma^3\Psi$
 for the region $I$ and $II$. 
\beqa
j^3_{I}&=&2kE\displaystyle\sum_s|C^s_+|^2(1-|R_s|^2) \nonumber, \\
j^3_{II}&=&2kE\displaystyle\sum_s|C^s_+|^2|T_s|^2(1-\frac{(-)^sm_s}{E}).
\label{j}
\eeqa
During this calculation the orthogonality condition 
$\bar{u}^s_\pm u^{s'}_\mp=\delta_{ss'}$ and 
$\bar{u}^s_\pm u^{s'}_\pm=0$ 
are useful. 
Here, from eq.(\ref{RT}) 
\beqa
|R_s|^2&=& \frac{2m_0(m_0-m_c)(1-(-)^s\frac{m_s}{E})}{F},
 \nonumber \\
|T_s|^2&=&\frac{4k^2}{F}
\label{rs}
\eeqa
with $F=2(2k^2+m_0(m_0-m_c))(1-(-)^sm_s/E)$.
We have used the relations
$E^2=k^2+m_0^2$ and $m_s^2+m_c^2=m_0^2$ repeatedly. 

So the reflection and the transmission coefficients
are given by
\beqa
R&=&\frac{\displaystyle\sum_s |C^s_+|^2 |R_s|^2}{\displaystyle\sum_s 
   |C^s_+|^2 }=
\frac{m_0(m_0-m_c)}{2k^2+m_0(m_0-m_c)}, \nonumber \\
T&=&\frac{\displaystyle\sum_s |C^s_+|^2 |T_s|^2 (1-(-)^s\frac{m_s}{E})}
   {\displaystyle\sum_s |C^s_+|^2 } 
=\frac{2k^2 }{2k^2+m_0(m_0-m_c)}. \nonumber \\
\label{results}
\eeqa
It is clear that the unitarity follows from eq.(\ref{results}), i.e., $R+T=1$.
We can get the coefficient for an antifermion 
by substituting $-E$, $-k$ and $C_+^{'s}$ (corresponding to $C^s_+$)
 instead of $E$, $k$ and $C^s_+$
 respectively. Since $R$ and $T$ are not changed
by these substitutions, we can say that
the fermion and the antifermion
 reflect at the phase wave in a CP conserving way.
It is contrary to the bubble wall case where
a position dependent phase of $\phi$ generally
induces a CP violating reflection.

Since $R$ and $T$ depend on $cos(\triangle\theta)$, there is
no difference between $\triangle\theta$ and $-\triangle\theta$.
It also means that $R$ and $T$ are
not dependent on the incoming direction of the fermion (left or right).

Now let us find approximate $R$ for $\triangle\theta\ll 1$.
In this case, $m_c\simeq m_0(1-\triangle\theta^2/2)$ and  $m_s\simeq m_0
\triangle\theta$. 
Gathering all terms up to $O(\triangle\theta^2)$
we find 
\beq
R \simeq\frac{m_0^2 \triangle \theta^2}{4k^2}.
\label{R}
\eeq
This implies that the particles with momentum $k\ll k_c\equiv
 m_0\triangle\theta/2$
will totally reflect at the phase wave and transfer
the momentum $2k$ to the wave.  

We also confirm this result (\ref{R}) by
solving eq.(\ref{dirac}) and eq.(\ref{KG})
in a second-order perturbation \cite{born}
when the width of the wave $d$ is finite.
There we checked 
that $R$ in eq.(\ref{R}) is the leading order term
at $d\rightarrow 0$ limit,
expanding $e^{i\theta}\simeq1+i\theta-\theta^2/2$ and 
finding solution up to the second order,

From eq.(\ref{results})
we find that the definition for  $k_c$ is also reasonable 
even for $\triangle\theta\simeq 1$.
Hence we will approximate $R(k)$ as $\theta(k_c-k)$.

When there is no viscosity, 
as the phase wave expands, it should lose
its gradient energy $(\partial_\mu\theta)^2$ and
its width $d$ becomes thicker and thicker, because,
contrary to the bubble wall case, there is
no net pressure from the vacuum energy difference
across the phase wave\cite{hawking}. 

 So it is possible that the particle scattering
may reduce the velocity of the wave and even stop  
it.

The viscosity by this reflection
could be calculated by the thermally averaged 
momentum transfer $\langle P\rangle$.

From the energy momentum relation $E^2=k_\bot^2+k_z^2+m^2_0$,
 where $k_\bot^2=k_x^2+k_y^2$, we find $k_zdk_z=EdE$.
In thermal equilibrium states,
  there are $n(E)d^3k/(2\pi)^3 |k_z/E|$ particles 
per unit wall area in unit time
providing a momentum transfer $2k_z$ \cite{arnold}.
Here $n(E)=(e^{E/T}+1)^{-1}$ is the Fermi distribution function
at the temperature $T$. 

Assuming $v>0$, we obtain
\beqa
\langle P \rangle=\int \frac{d^2k_\bot}{(2\pi)^3} \{
\int^{\infty}_{0} dk_z 2k_zR(k_z)\frac{k_z}{E} n[\gamma(E+vk_z)] \nonumber \\
+\int^{0}_{-\infty} dk_z 2k_z R(k_z)\frac{-k_z}{E} n[\gamma(E+vk_z)] \}.
\eeqa
where $d^2k_\bot=dk_xdk_y$.
The first term corresponds to the pressure by the fermions with
$k_z>0$, while the second term corresponds to that with $k_z<0$.
Changing integration variable from $k_z$ to $-k_z$ in the second term
we obtain 
\beq
\langle P \rangle =\int \frac{d^2k_\bot}{(2\pi)^3} 
\int^{\infty}_{0} dk_z 2\theta(k_c-k_z)\frac{k_z^2}{E}\{ n[\gamma(E+vk_z)] 
-n[\gamma(E-vk_z)] \}
\label{P}
\eeq
Fig. 2. displays a numerical result of this integration.
$\langle P \rangle $ increases linearly as $v$ increases, and then goes
to zero as $v$ approaches to $1$.

Now we assume that the velocity of the wave is nonrelativistic
($v\ll1$).
Then $n(\gamma(E+vk_z))\simeq n(E)+v k_z\frac{\partial n(E)}
{\partial E}=n(E)+v\beta k_zn(E)(n(E)-1)$.
Changing integration variables 
from $d^3k$ to $2\pi E dE dk_z $ and assuming $\beta E\gg 1$ give
\beqa
\langle P \rangle &=&
-v\frac{\beta}{\pi^2} \int^\infty_{m_0} dE e^{-\beta E}
\int^{\sqrt{E^2-m_0^2}}_0dk_z \theta(k_c-k_z)k_z^3
 \nonumber\\
&=&-v\frac{\beta}{\pi^2} \{ \int^{E_c}_{m_0} dE e^{-\beta E}
\int^{\sqrt{E^2-m_0^2}}_0dk_z \theta(k_c-k_z)k_z^3 \nonumber\\
&+& \int^{\infty}_{E_c} dE e^{-\beta E}
[\int^{k_c}_0dk_z \theta(k_c-k_z)k_z^3
+\int^{\sqrt{E^2-m_0^2}}_{k_c} dk_z \theta(k_c-k_z)k_z^3]\},\nonumber\\
\eeqa
where $E_c^2\equiv m_0^2+k_c^2=m^2_0(1+\triangle\theta^2/4)$ 
and we used  the approximation 
$n(E)\simeq exp(-\beta E)$.
We have separated the integration regions to
drop the third term.

Then,
\beqa
\langle P \rangle
&\simeq& 
-v\frac{\beta}{4\pi^2}\int^{E_c}_{m_0} dE e^{-\beta E_c} (E^2-m_0^2)^2
-v\frac{ k_c^4}{4\pi^2}e^{-\beta E_c} \nonumber \\
&\simeq& 
-v\frac{ k_c^4}{4\pi^2}e^{-\beta E_c},
\label{P2}
\eeqa 
because the first term is of order $(\triangle\theta)^6$,
while the second term is of order  $(\triangle\theta)^4$.

For $\theta=\theta(r-vt)$
 the Newtonian equation for
$r$  is given by\cite{equation}
\beq
\sigma\frac{d^2r}{dt^2}=-\frac{2\sigma}{r}+
\langle P \rangle 
\eeq
where $v$ is now $dr/dt$.
This equation can be derived by considering the force acting on
the unit spherical area of the wave with the `radius' $r$ and `mass' density
$\sigma$.
The surface energy density $\sigma$ is related
to the energy momentum tensor by
\beq
\sigma=\int dz T^0_0=\int dz \frac{1}{2} (\partial_z\phi)^2
\simeq \frac{\rho^2}{d}.
\eeq

Hence for nonrelativistic and planar approximation($r\rightarrow\infty$)
\beq
\sigma\frac{d^2r}{dt^2}=\sigma \frac{dr}{dt} v\simeq
-v\frac{ k_c^4}{4\pi^2}e^{-\beta E_c},
\eeq
which has a solution 
\beq
v= v_0 exp[-\frac{k_c^4}
{4\pi^2\sigma}exp(-\beta E_c )t],
\eeq
where $v_0$ is an initial velocity.

Hence  the phase wave stops in a time scale 
$t_{stop}\equiv \frac{4\pi^2\sigma}{k_c^4}e^{\beta E_c}$
and the wave travels the distance 
\beq
\triangle r \simeq v_0 t_{stop} [1-exp(- t/t_{stop})]
\eeq
during $t$.

The bubble nucleation rate per unit time and volume is
$\Gamma\simeq T^4 e^{-A}$ and $ A\simeq 4 ln M_P/T$\cite{size}.
The average bubble distance and the
bubble percolation time scale is $t_{bub}\sim\beta^{-1}$,
where  $\beta=\gamma A H$. Here the potential form
dependent factor $\gamma=dlnA/dlnT$ is  generally
$O(1)\sim O(10^2)$ \cite{adams}. 

For example, if $T\simeq 10^2$  $GeV$, $A\simeq 10^2$ and
$t_{bub}\simeq 10^2 H^{-1}\simeq 10^2 M_P/T^2$.
On the other hand,
 if all the coupling constants are not too smaller than
unity, then  $m_0\sim T$,  and $t_{stop}\sim T^{-1}e^{\beta E_c}$.
Hence, for $\beta$ not too smaller than $E_c$, 
$t_{stop}$ should be shorter than $t_{bub}$.  
Generally if $T \ll e^{-\beta E_c} M_P$, the phase
wave may stop before it reaches the other side
of the bubble wall.

If  these stopped phase waves 
persist long time, their energy density may give rise to a problem
similar to the domain wall problem\cite{kibble}.
Even if some ignored mechanisms ( for example surface tension)
 make the wave disappear,
they take a role only within the horizon at that time and
hence the problem could not removed completely.
 So in this view point, it might be preferable
to have a non-degenerate unique vacuum for $\langle \phi \rangle$.

When the Yukawa coupling $h$ is  complex,
defining $m_c\equiv m_0(Re(h)cos(\triangle\theta)
-Im(h)sin(\triangle\theta))$
and $m_s\equiv m_0(Re(h)sin(\triangle\theta)
+Im(h)cos(\triangle\theta))$
gives the same result in eq.(\ref{results}).

If there is another scalar field $\chi$ interacting with $\phi$
through $(\lambda/2)\lambda|\phi|^2|\chi|^2$, $\chi$ gets
a constant mass $\lambda^{1/2} \rho$ and satisfies Klein-Gordon
equation. Hence the scalar field $\chi$ does not
reflect at the phase wave.

The origin of fermion scattering at the phase wave is
a position dependent phase change of its mass, while
that of the bubble wall is  a change of absolute mass and,
therefore, a momentum change across the wall.  

Our results may be also applied to detection
of a spatial phase change and its velocity $v$,
if any, in the universe or a condensed matter. 
For example, by using eq.(\ref{R}) and
 observing the difference between 
the two $R$'s for electron beams 
with the opposite incoming
directions. (say, from region I and II in fig.1., respectively.)
Each electron beams should have different $k$ 
in the phase wave rest frame moving with $v$, and have different $R$.

The results, so far, are for a global symmetry without gauge fields.
Including gauge fields may give more complicated and interesting results. 

In summary, we calculated the reflection coefficient
of the fermion scattering off the phase wave,
which could be a first step toward a comprehensive
understanding of the phase wave  dynamics in the hot plasma.
As an example, we investigated the damping  and the velocity
of the wave due to this scattering,
and found that this damping could even stop the wave in
some cases.

\vskip 1cm
 One of the authors (Lee) is thankful to H.M. Kwon, J.D. Kim, Y.D.
 Han and H.C. Kim for helpful discussions. 
 This work was supported in part by KOSEF.

\newpage

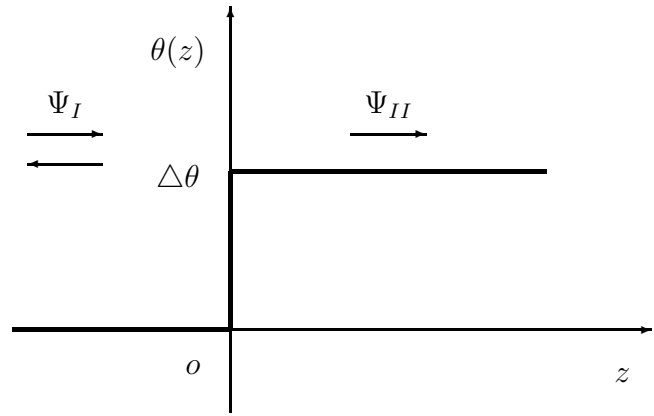
{\bf FIG.1.}\\
$\theta(z)$ in the phase wave rest frame. 
$\Psi_I$ consists of the incoming and the reflected waves
and $\Psi_{II}$ is the transmitted wave.

\vspace{2cm}

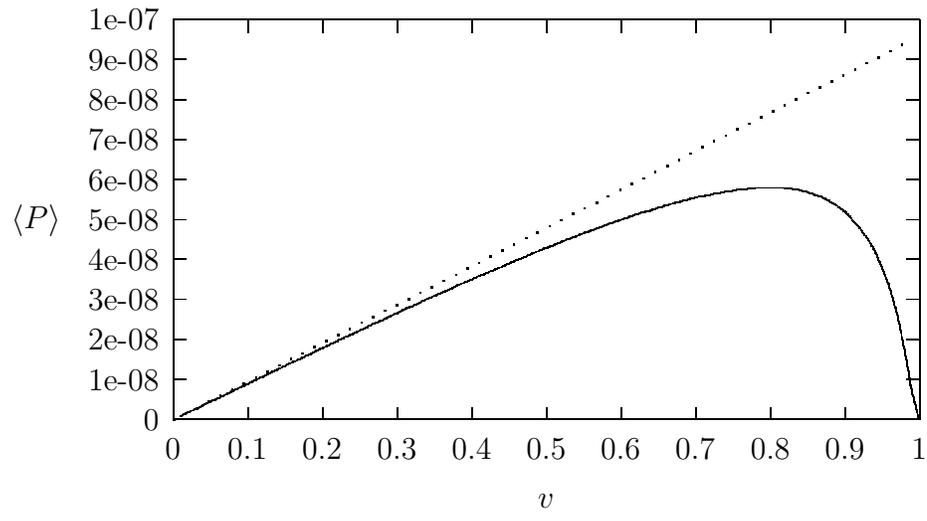
{\bf FIG.2.}\\
The damping pressure $\langle P \rangle $ in units of $m_0^4$
 versus the velocity of the phase wave
 $v$ for $\triangle\theta=0.1$ and $T=2 m_0$. 
The solid curve shows the results of eq.(\ref{P})
and the dotted line shows the results of eq.(\ref{P2}).

\newpage  

\unitlength=1.00mm
\linethickness{0.8pt}
\begin{picture}(113.00,123.00)
\put(28.00,80.00){\vector(1,0){85.00}}
\put(57.00,69.00){\vector(0,1){54.00}}
\put(30.00,106.00){\vector(1,0){10.00}}
\put(40.00,102.00){\vector(-1,0){10.00}}
\put(73.00,106.00){\vector(1,0){10.00}}
\linethickness{1.5pt}
\put(108.00,74.00){\makebox(0,0)[lc]{$z$}}
\put(50.00,117.00){\makebox(0,0)[cc]{$\theta (z)$}}
\put(50.00,100.00){\makebox(0,0)[cc]{$\triangle\theta$}}

\put(35.00,110.00){\makebox(0,0)[cc]{$\Psi_I$}}
\put(78.00,110.00){\makebox(0,0)[cc]{$\Psi_{II}$}}

\put(28.00,80.00){\line(1,0){29.00}}
\put(52.00,75.00){\makebox(0,0)[cc]{$o$}}
\put(57.00,101.00){\line(1,0){42.00}}
\put(99.00,101.00){\line(0,0){0.00}}
\put(99.00,101.00){\line(0,0){0.00}}
\put(57.00,101.00){\line(0,-1){21.00}}
\put(50,-20){Fig. 1.}
\end{picture}

\newpage

\setlength{\unitlength}{0.240900pt}
\ifx\plotpoint\undefined\newsavebox{\plotpoint}\fi
\sbox{\plotpoint}{\rule[-0.175pt]{0.350pt}{0.350pt}}%
\begin{picture}(1500,900)(0,0)
\sbox{\plotpoint}{\rule[-0.175pt]{0.350pt}{0.350pt}}%
\put(264,158){\rule[-0.175pt]{282.335pt}{0.350pt}}
\put(264,158){\rule[-0.175pt]{0.350pt}{151.526pt}}
\put(264,158){\rule[-0.175pt]{4.818pt}{0.350pt}}
\put(242,158){\makebox(0,0)[r]{0}}
\put(1416,158){\rule[-0.175pt]{4.818pt}{0.350pt}}
\put(264,221){\rule[-0.175pt]{4.818pt}{0.350pt}}
\put(242,221){\makebox(0,0)[r]{1e-08}}
\put(1416,221){\rule[-0.175pt]{4.818pt}{0.350pt}}
\put(264,284){\rule[-0.175pt]{4.818pt}{0.350pt}}
\put(242,284){\makebox(0,0)[r]{2e-08}}
\put(1416,284){\rule[-0.175pt]{4.818pt}{0.350pt}}
\put(264,347){\rule[-0.175pt]{4.818pt}{0.350pt}}
\put(242,347){\makebox(0,0)[r]{3e-08}}
\put(1416,347){\rule[-0.175pt]{4.818pt}{0.350pt}}
\put(264,410){\rule[-0.175pt]{4.818pt}{0.350pt}}
\put(242,410){\makebox(0,0)[r]{4e-08}}
\put(1416,410){\rule[-0.175pt]{4.818pt}{0.350pt}}
\put(264,472){\rule[-0.175pt]{4.818pt}{0.350pt}}
\put(242,472){\makebox(0,0)[r]{5e-08}}
\put(1416,472){\rule[-0.175pt]{4.818pt}{0.350pt}}
\put(264,535){\rule[-0.175pt]{4.818pt}{0.350pt}}
\put(242,535){\makebox(0,0)[r]{6e-08}}
\put(1416,535){\rule[-0.175pt]{4.818pt}{0.350pt}}
\put(264,598){\rule[-0.175pt]{4.818pt}{0.350pt}}
\put(242,598){\makebox(0,0)[r]{7e-08}}
\put(1416,598){\rule[-0.175pt]{4.818pt}{0.350pt}}
\put(264,661){\rule[-0.175pt]{4.818pt}{0.350pt}}
\put(242,661){\makebox(0,0)[r]{8e-08}}
\put(1416,661){\rule[-0.175pt]{4.818pt}{0.350pt}}
\put(264,724){\rule[-0.175pt]{4.818pt}{0.350pt}}
\put(242,724){\makebox(0,0)[r]{9e-08}}
\put(1416,724){\rule[-0.175pt]{4.818pt}{0.350pt}}
\put(264,787){\rule[-0.175pt]{4.818pt}{0.350pt}}
\put(242,787){\makebox(0,0)[r]{1e-07}}
\put(1416,787){\rule[-0.175pt]{4.818pt}{0.350pt}}
\put(264,158){\rule[-0.175pt]{0.350pt}{4.818pt}}
\put(264,113){\makebox(0,0){0}}
\put(264,767){\rule[-0.175pt]{0.350pt}{4.818pt}}
\put(381,158){\rule[-0.175pt]{0.350pt}{4.818pt}}
\put(381,113){\makebox(0,0){0.1}}
\put(381,767){\rule[-0.175pt]{0.350pt}{4.818pt}}
\put(498,158){\rule[-0.175pt]{0.350pt}{4.818pt}}
\put(498,113){\makebox(0,0){0.2}}
\put(498,767){\rule[-0.175pt]{0.350pt}{4.818pt}}
\put(616,158){\rule[-0.175pt]{0.350pt}{4.818pt}}
\put(616,113){\makebox(0,0){0.3}}
\put(616,767){\rule[-0.175pt]{0.350pt}{4.818pt}}
\put(733,158){\rule[-0.175pt]{0.350pt}{4.818pt}}
\put(733,113){\makebox(0,0){0.4}}
\put(733,767){\rule[-0.175pt]{0.350pt}{4.818pt}}
\put(850,158){\rule[-0.175pt]{0.350pt}{4.818pt}}
\put(850,113){\makebox(0,0){0.5}}
\put(850,767){\rule[-0.175pt]{0.350pt}{4.818pt}}
\put(967,158){\rule[-0.175pt]{0.350pt}{4.818pt}}
\put(967,113){\makebox(0,0){0.6}}
\put(967,767){\rule[-0.175pt]{0.350pt}{4.818pt}}
\put(1084,158){\rule[-0.175pt]{0.350pt}{4.818pt}}
\put(1084,113){\makebox(0,0){0.7}}
\put(1084,767){\rule[-0.175pt]{0.350pt}{4.818pt}}
\put(1202,158){\rule[-0.175pt]{0.350pt}{4.818pt}}
\put(1202,113){\makebox(0,0){0.8}}
\put(1202,767){\rule[-0.175pt]{0.350pt}{4.818pt}}
\put(1319,158){\rule[-0.175pt]{0.350pt}{4.818pt}}
\put(1319,113){\makebox(0,0){0.9}}
\put(1319,767){\rule[-0.175pt]{0.350pt}{4.818pt}}
\put(1436,158){\rule[-0.175pt]{0.350pt}{4.818pt}}
\put(1436,113){\makebox(0,0){1}}
\put(1436,767){\rule[-0.175pt]{0.350pt}{4.818pt}}
\put(264,158){\rule[-0.175pt]{282.335pt}{0.350pt}}
\put(1436,158){\rule[-0.175pt]{0.350pt}{151.526pt}}
\put(264,787){\rule[-0.175pt]{282.335pt}{0.350pt}}

\put(10,472){\makebox(0,0)[l]{\shortstack{$\langle P \rangle $}}}
\put(850,30){\makebox(0,0){$v$}}
\put(264,158){\rule[-0.175pt]{0.350pt}{151.526pt}}
\put(275,163){\usebox{\plotpoint}}
\put(275,163){\rule[-0.175pt]{0.442pt}{0.350pt}}
\put(276,164){\rule[-0.175pt]{0.442pt}{0.350pt}}
\put(278,165){\rule[-0.175pt]{0.442pt}{0.350pt}}
\put(280,166){\rule[-0.175pt]{0.442pt}{0.350pt}}
\put(282,167){\rule[-0.175pt]{0.442pt}{0.350pt}}
\put(284,168){\rule[-0.175pt]{0.442pt}{0.350pt}}
\put(286,169){\rule[-0.175pt]{0.578pt}{0.350pt}}
\put(288,170){\rule[-0.175pt]{0.578pt}{0.350pt}}
\put(290,171){\rule[-0.175pt]{0.578pt}{0.350pt}}
\put(293,172){\rule[-0.175pt]{0.578pt}{0.350pt}}
\put(295,173){\rule[-0.175pt]{0.578pt}{0.350pt}}
\put(297,174){\rule[-0.175pt]{0.482pt}{0.350pt}}
\put(300,175){\rule[-0.175pt]{0.482pt}{0.350pt}}
\put(302,176){\rule[-0.175pt]{0.482pt}{0.350pt}}
\put(304,177){\rule[-0.175pt]{0.482pt}{0.350pt}}
\put(306,178){\rule[-0.175pt]{0.482pt}{0.350pt}}
\put(308,179){\rule[-0.175pt]{0.482pt}{0.350pt}}
\put(310,180){\rule[-0.175pt]{0.442pt}{0.350pt}}
\put(311,181){\rule[-0.175pt]{0.442pt}{0.350pt}}
\put(313,182){\rule[-0.175pt]{0.442pt}{0.350pt}}
\put(315,183){\rule[-0.175pt]{0.442pt}{0.350pt}}
\put(317,184){\rule[-0.175pt]{0.442pt}{0.350pt}}
\put(319,185){\rule[-0.175pt]{0.442pt}{0.350pt}}
\put(321,186){\rule[-0.175pt]{0.578pt}{0.350pt}}
\put(323,187){\rule[-0.175pt]{0.578pt}{0.350pt}}
\put(325,188){\rule[-0.175pt]{0.578pt}{0.350pt}}
\put(328,189){\rule[-0.175pt]{0.578pt}{0.350pt}}
\put(330,190){\rule[-0.175pt]{0.578pt}{0.350pt}}
\put(332,191){\rule[-0.175pt]{0.482pt}{0.350pt}}
\put(335,192){\rule[-0.175pt]{0.482pt}{0.350pt}}
\put(337,193){\rule[-0.175pt]{0.482pt}{0.350pt}}
\put(339,194){\rule[-0.175pt]{0.482pt}{0.350pt}}
\put(341,195){\rule[-0.175pt]{0.482pt}{0.350pt}}
\put(343,196){\rule[-0.175pt]{0.482pt}{0.350pt}}
\put(345,197){\rule[-0.175pt]{0.482pt}{0.350pt}}
\put(347,198){\rule[-0.175pt]{0.482pt}{0.350pt}}
\put(349,199){\rule[-0.175pt]{0.482pt}{0.350pt}}
\put(351,200){\rule[-0.175pt]{0.482pt}{0.350pt}}
\put(353,201){\rule[-0.175pt]{0.482pt}{0.350pt}}
\put(355,202){\rule[-0.175pt]{0.482pt}{0.350pt}}
\put(357,203){\rule[-0.175pt]{0.530pt}{0.350pt}}
\put(359,204){\rule[-0.175pt]{0.530pt}{0.350pt}}
\put(361,205){\rule[-0.175pt]{0.530pt}{0.350pt}}
\put(363,206){\rule[-0.175pt]{0.530pt}{0.350pt}}
\put(365,207){\rule[-0.175pt]{0.530pt}{0.350pt}}
\put(368,208){\rule[-0.175pt]{0.482pt}{0.350pt}}
\put(370,209){\rule[-0.175pt]{0.482pt}{0.350pt}}
\put(372,210){\rule[-0.175pt]{0.482pt}{0.350pt}}
\put(374,211){\rule[-0.175pt]{0.482pt}{0.350pt}}
\put(376,212){\rule[-0.175pt]{0.482pt}{0.350pt}}
\put(378,213){\rule[-0.175pt]{0.482pt}{0.350pt}}
\put(380,214){\rule[-0.175pt]{0.482pt}{0.350pt}}
\put(382,215){\rule[-0.175pt]{0.482pt}{0.350pt}}
\put(384,216){\rule[-0.175pt]{0.482pt}{0.350pt}}
\put(386,217){\rule[-0.175pt]{0.482pt}{0.350pt}}
\put(388,218){\rule[-0.175pt]{0.482pt}{0.350pt}}
\put(390,219){\rule[-0.175pt]{0.482pt}{0.350pt}}
\put(392,220){\rule[-0.175pt]{0.530pt}{0.350pt}}
\put(394,221){\rule[-0.175pt]{0.530pt}{0.350pt}}
\put(396,222){\rule[-0.175pt]{0.530pt}{0.350pt}}
\put(398,223){\rule[-0.175pt]{0.530pt}{0.350pt}}
\put(400,224){\rule[-0.175pt]{0.530pt}{0.350pt}}
\put(403,225){\rule[-0.175pt]{0.482pt}{0.350pt}}
\put(405,226){\rule[-0.175pt]{0.482pt}{0.350pt}}
\put(407,227){\rule[-0.175pt]{0.482pt}{0.350pt}}
\put(409,228){\rule[-0.175pt]{0.482pt}{0.350pt}}
\put(411,229){\rule[-0.175pt]{0.482pt}{0.350pt}}
\put(413,230){\rule[-0.175pt]{0.482pt}{0.350pt}}
\put(415,231){\rule[-0.175pt]{0.482pt}{0.350pt}}
\put(417,232){\rule[-0.175pt]{0.482pt}{0.350pt}}
\put(419,233){\rule[-0.175pt]{0.482pt}{0.350pt}}
\put(421,234){\rule[-0.175pt]{0.482pt}{0.350pt}}
\put(423,235){\rule[-0.175pt]{0.482pt}{0.350pt}}
\put(425,236){\rule[-0.175pt]{0.482pt}{0.350pt}}
\put(427,237){\rule[-0.175pt]{0.578pt}{0.350pt}}
\put(429,238){\rule[-0.175pt]{0.578pt}{0.350pt}}
\put(431,239){\rule[-0.175pt]{0.578pt}{0.350pt}}
\put(434,240){\rule[-0.175pt]{0.578pt}{0.350pt}}
\put(436,241){\rule[-0.175pt]{0.578pt}{0.350pt}}
\put(438,242){\rule[-0.175pt]{0.442pt}{0.350pt}}
\put(440,243){\rule[-0.175pt]{0.442pt}{0.350pt}}
\put(442,244){\rule[-0.175pt]{0.442pt}{0.350pt}}
\put(444,245){\rule[-0.175pt]{0.442pt}{0.350pt}}
\put(446,246){\rule[-0.175pt]{0.442pt}{0.350pt}}
\put(448,247){\rule[-0.175pt]{0.442pt}{0.350pt}}
\put(450,248){\rule[-0.175pt]{0.578pt}{0.350pt}}
\put(452,249){\rule[-0.175pt]{0.578pt}{0.350pt}}
\put(454,250){\rule[-0.175pt]{0.578pt}{0.350pt}}
\put(457,251){\rule[-0.175pt]{0.578pt}{0.350pt}}
\put(459,252){\rule[-0.175pt]{0.578pt}{0.350pt}}
\put(461,253){\rule[-0.175pt]{0.482pt}{0.350pt}}
\put(464,254){\rule[-0.175pt]{0.482pt}{0.350pt}}
\put(466,255){\rule[-0.175pt]{0.482pt}{0.350pt}}
\put(468,256){\rule[-0.175pt]{0.482pt}{0.350pt}}
\put(470,257){\rule[-0.175pt]{0.482pt}{0.350pt}}
\put(472,258){\rule[-0.175pt]{0.482pt}{0.350pt}}
\put(474,259){\rule[-0.175pt]{0.482pt}{0.350pt}}
\put(476,260){\rule[-0.175pt]{0.482pt}{0.350pt}}
\put(478,261){\rule[-0.175pt]{0.482pt}{0.350pt}}
\put(480,262){\rule[-0.175pt]{0.482pt}{0.350pt}}
\put(482,263){\rule[-0.175pt]{0.482pt}{0.350pt}}
\put(484,264){\rule[-0.175pt]{0.482pt}{0.350pt}}
\put(486,265){\rule[-0.175pt]{0.530pt}{0.350pt}}
\put(488,266){\rule[-0.175pt]{0.530pt}{0.350pt}}
\put(490,267){\rule[-0.175pt]{0.530pt}{0.350pt}}
\put(492,268){\rule[-0.175pt]{0.530pt}{0.350pt}}
\put(494,269){\rule[-0.175pt]{0.530pt}{0.350pt}}
\put(497,270){\rule[-0.175pt]{0.482pt}{0.350pt}}
\put(499,271){\rule[-0.175pt]{0.482pt}{0.350pt}}
\put(501,272){\rule[-0.175pt]{0.482pt}{0.350pt}}
\put(503,273){\rule[-0.175pt]{0.482pt}{0.350pt}}
\put(505,274){\rule[-0.175pt]{0.482pt}{0.350pt}}
\put(507,275){\rule[-0.175pt]{0.482pt}{0.350pt}}
\put(509,276){\rule[-0.175pt]{0.578pt}{0.350pt}}
\put(511,277){\rule[-0.175pt]{0.578pt}{0.350pt}}
\put(513,278){\rule[-0.175pt]{0.578pt}{0.350pt}}
\put(516,279){\rule[-0.175pt]{0.578pt}{0.350pt}}
\put(518,280){\rule[-0.175pt]{0.578pt}{0.350pt}}
\put(521,281){\rule[-0.175pt]{0.442pt}{0.350pt}}
\put(522,282){\rule[-0.175pt]{0.442pt}{0.350pt}}
\put(524,283){\rule[-0.175pt]{0.442pt}{0.350pt}}
\put(526,284){\rule[-0.175pt]{0.442pt}{0.350pt}}
\put(528,285){\rule[-0.175pt]{0.442pt}{0.350pt}}
\put(530,286){\rule[-0.175pt]{0.442pt}{0.350pt}}
\put(531,287){\rule[-0.175pt]{0.578pt}{0.350pt}}
\put(534,288){\rule[-0.175pt]{0.578pt}{0.350pt}}
\put(536,289){\rule[-0.175pt]{0.578pt}{0.350pt}}
\put(539,290){\rule[-0.175pt]{0.578pt}{0.350pt}}
\put(541,291){\rule[-0.175pt]{0.578pt}{0.350pt}}
\put(544,292){\rule[-0.175pt]{0.482pt}{0.350pt}}
\put(546,293){\rule[-0.175pt]{0.482pt}{0.350pt}}
\put(548,294){\rule[-0.175pt]{0.482pt}{0.350pt}}
\put(550,295){\rule[-0.175pt]{0.482pt}{0.350pt}}
\put(552,296){\rule[-0.175pt]{0.482pt}{0.350pt}}
\put(554,297){\rule[-0.175pt]{0.482pt}{0.350pt}}
\put(556,298){\rule[-0.175pt]{0.578pt}{0.350pt}}
\put(558,299){\rule[-0.175pt]{0.578pt}{0.350pt}}
\put(560,300){\rule[-0.175pt]{0.578pt}{0.350pt}}
\put(563,301){\rule[-0.175pt]{0.578pt}{0.350pt}}
\put(565,302){\rule[-0.175pt]{0.578pt}{0.350pt}}
\put(568,303){\rule[-0.175pt]{0.442pt}{0.350pt}}
\put(569,304){\rule[-0.175pt]{0.442pt}{0.350pt}}
\put(571,305){\rule[-0.175pt]{0.442pt}{0.350pt}}
\put(573,306){\rule[-0.175pt]{0.442pt}{0.350pt}}
\put(575,307){\rule[-0.175pt]{0.442pt}{0.350pt}}
\put(577,308){\rule[-0.175pt]{0.442pt}{0.350pt}}
\put(578,309){\rule[-0.175pt]{0.578pt}{0.350pt}}
\put(581,310){\rule[-0.175pt]{0.578pt}{0.350pt}}
\put(583,311){\rule[-0.175pt]{0.578pt}{0.350pt}}
\put(586,312){\rule[-0.175pt]{0.578pt}{0.350pt}}
\put(588,313){\rule[-0.175pt]{0.578pt}{0.350pt}}
\put(591,314){\rule[-0.175pt]{0.482pt}{0.350pt}}
\put(593,315){\rule[-0.175pt]{0.482pt}{0.350pt}}
\put(595,316){\rule[-0.175pt]{0.482pt}{0.350pt}}
\put(597,317){\rule[-0.175pt]{0.482pt}{0.350pt}}
\put(599,318){\rule[-0.175pt]{0.482pt}{0.350pt}}
\put(601,319){\rule[-0.175pt]{0.482pt}{0.350pt}}
\put(603,320){\rule[-0.175pt]{0.530pt}{0.350pt}}
\put(605,321){\rule[-0.175pt]{0.530pt}{0.350pt}}
\put(607,322){\rule[-0.175pt]{0.530pt}{0.350pt}}
\put(609,323){\rule[-0.175pt]{0.530pt}{0.350pt}}
\put(611,324){\rule[-0.175pt]{0.530pt}{0.350pt}}
\put(614,325){\rule[-0.175pt]{0.482pt}{0.350pt}}
\put(616,326){\rule[-0.175pt]{0.482pt}{0.350pt}}
\put(618,327){\rule[-0.175pt]{0.482pt}{0.350pt}}
\put(620,328){\rule[-0.175pt]{0.482pt}{0.350pt}}
\put(622,329){\rule[-0.175pt]{0.482pt}{0.350pt}}
\put(624,330){\rule[-0.175pt]{0.482pt}{0.350pt}}
\put(626,331){\rule[-0.175pt]{0.578pt}{0.350pt}}
\put(628,332){\rule[-0.175pt]{0.578pt}{0.350pt}}
\put(630,333){\rule[-0.175pt]{0.578pt}{0.350pt}}
\put(633,334){\rule[-0.175pt]{0.578pt}{0.350pt}}
\put(635,335){\rule[-0.175pt]{0.578pt}{0.350pt}}
\put(638,336){\rule[-0.175pt]{0.578pt}{0.350pt}}
\put(640,337){\rule[-0.175pt]{0.578pt}{0.350pt}}
\put(642,338){\rule[-0.175pt]{0.578pt}{0.350pt}}
\put(645,339){\rule[-0.175pt]{0.578pt}{0.350pt}}
\put(647,340){\rule[-0.175pt]{0.578pt}{0.350pt}}
\put(650,341){\rule[-0.175pt]{0.442pt}{0.350pt}}
\put(651,342){\rule[-0.175pt]{0.442pt}{0.350pt}}
\put(653,343){\rule[-0.175pt]{0.442pt}{0.350pt}}
\put(655,344){\rule[-0.175pt]{0.442pt}{0.350pt}}
\put(657,345){\rule[-0.175pt]{0.442pt}{0.350pt}}
\put(659,346){\rule[-0.175pt]{0.442pt}{0.350pt}}
\put(660,347){\rule[-0.175pt]{0.578pt}{0.350pt}}
\put(663,348){\rule[-0.175pt]{0.578pt}{0.350pt}}
\put(665,349){\rule[-0.175pt]{0.578pt}{0.350pt}}
\put(668,350){\rule[-0.175pt]{0.578pt}{0.350pt}}
\put(670,351){\rule[-0.175pt]{0.578pt}{0.350pt}}
\put(673,352){\rule[-0.175pt]{0.578pt}{0.350pt}}
\put(675,353){\rule[-0.175pt]{0.578pt}{0.350pt}}
\put(677,354){\rule[-0.175pt]{0.578pt}{0.350pt}}
\put(680,355){\rule[-0.175pt]{0.578pt}{0.350pt}}
\put(682,356){\rule[-0.175pt]{0.578pt}{0.350pt}}
\put(685,357){\rule[-0.175pt]{0.530pt}{0.350pt}}
\put(687,358){\rule[-0.175pt]{0.530pt}{0.350pt}}
\put(689,359){\rule[-0.175pt]{0.530pt}{0.350pt}}
\put(691,360){\rule[-0.175pt]{0.530pt}{0.350pt}}
\put(693,361){\rule[-0.175pt]{0.530pt}{0.350pt}}
\put(696,362){\rule[-0.175pt]{0.482pt}{0.350pt}}
\put(698,363){\rule[-0.175pt]{0.482pt}{0.350pt}}
\put(700,364){\rule[-0.175pt]{0.482pt}{0.350pt}}
\put(702,365){\rule[-0.175pt]{0.482pt}{0.350pt}}
\put(704,366){\rule[-0.175pt]{0.482pt}{0.350pt}}
\put(706,367){\rule[-0.175pt]{0.482pt}{0.350pt}}
\put(708,368){\rule[-0.175pt]{0.578pt}{0.350pt}}
\put(710,369){\rule[-0.175pt]{0.578pt}{0.350pt}}
\put(712,370){\rule[-0.175pt]{0.578pt}{0.350pt}}
\put(715,371){\rule[-0.175pt]{0.578pt}{0.350pt}}
\put(717,372){\rule[-0.175pt]{0.578pt}{0.350pt}}
\put(720,373){\rule[-0.175pt]{0.578pt}{0.350pt}}
\put(722,374){\rule[-0.175pt]{0.578pt}{0.350pt}}
\put(724,375){\rule[-0.175pt]{0.578pt}{0.350pt}}
\put(727,376){\rule[-0.175pt]{0.578pt}{0.350pt}}
\put(729,377){\rule[-0.175pt]{0.578pt}{0.350pt}}
\put(732,378){\rule[-0.175pt]{0.530pt}{0.350pt}}
\put(734,379){\rule[-0.175pt]{0.530pt}{0.350pt}}
\put(736,380){\rule[-0.175pt]{0.530pt}{0.350pt}}
\put(738,381){\rule[-0.175pt]{0.530pt}{0.350pt}}
\put(740,382){\rule[-0.175pt]{0.530pt}{0.350pt}}
\put(743,383){\rule[-0.175pt]{0.578pt}{0.350pt}}
\put(745,384){\rule[-0.175pt]{0.578pt}{0.350pt}}
\put(747,385){\rule[-0.175pt]{0.578pt}{0.350pt}}
\put(750,386){\rule[-0.175pt]{0.578pt}{0.350pt}}
\put(752,387){\rule[-0.175pt]{0.578pt}{0.350pt}}
\put(755,388){\rule[-0.175pt]{0.578pt}{0.350pt}}
\put(757,389){\rule[-0.175pt]{0.578pt}{0.350pt}}
\put(759,390){\rule[-0.175pt]{0.578pt}{0.350pt}}
\put(762,391){\rule[-0.175pt]{0.578pt}{0.350pt}}
\put(764,392){\rule[-0.175pt]{0.578pt}{0.350pt}}
\put(767,393){\rule[-0.175pt]{0.578pt}{0.350pt}}
\put(769,394){\rule[-0.175pt]{0.578pt}{0.350pt}}
\put(771,395){\rule[-0.175pt]{0.578pt}{0.350pt}}
\put(774,396){\rule[-0.175pt]{0.578pt}{0.350pt}}
\put(776,397){\rule[-0.175pt]{0.578pt}{0.350pt}}
\put(779,398){\rule[-0.175pt]{0.530pt}{0.350pt}}
\put(781,399){\rule[-0.175pt]{0.530pt}{0.350pt}}
\put(783,400){\rule[-0.175pt]{0.530pt}{0.350pt}}
\put(785,401){\rule[-0.175pt]{0.530pt}{0.350pt}}
\put(787,402){\rule[-0.175pt]{0.530pt}{0.350pt}}
\put(790,403){\rule[-0.175pt]{0.578pt}{0.350pt}}
\put(792,404){\rule[-0.175pt]{0.578pt}{0.350pt}}
\put(794,405){\rule[-0.175pt]{0.578pt}{0.350pt}}
\put(797,406){\rule[-0.175pt]{0.578pt}{0.350pt}}
\put(799,407){\rule[-0.175pt]{0.578pt}{0.350pt}}
\put(802,408){\rule[-0.175pt]{0.578pt}{0.350pt}}
\put(804,409){\rule[-0.175pt]{0.578pt}{0.350pt}}
\put(806,410){\rule[-0.175pt]{0.578pt}{0.350pt}}
\put(809,411){\rule[-0.175pt]{0.578pt}{0.350pt}}
\put(811,412){\rule[-0.175pt]{0.578pt}{0.350pt}}
\put(814,413){\rule[-0.175pt]{0.530pt}{0.350pt}}
\put(816,414){\rule[-0.175pt]{0.530pt}{0.350pt}}
\put(818,415){\rule[-0.175pt]{0.530pt}{0.350pt}}
\put(820,416){\rule[-0.175pt]{0.530pt}{0.350pt}}
\put(822,417){\rule[-0.175pt]{0.530pt}{0.350pt}}
\put(825,418){\rule[-0.175pt]{0.578pt}{0.350pt}}
\put(827,419){\rule[-0.175pt]{0.578pt}{0.350pt}}
\put(829,420){\rule[-0.175pt]{0.578pt}{0.350pt}}
\put(832,421){\rule[-0.175pt]{0.578pt}{0.350pt}}
\put(834,422){\rule[-0.175pt]{0.578pt}{0.350pt}}
\put(837,423){\rule[-0.175pt]{0.578pt}{0.350pt}}
\put(839,424){\rule[-0.175pt]{0.578pt}{0.350pt}}
\put(841,425){\rule[-0.175pt]{0.578pt}{0.350pt}}
\put(844,426){\rule[-0.175pt]{0.578pt}{0.350pt}}
\put(846,427){\rule[-0.175pt]{0.578pt}{0.350pt}}
\put(849,428){\rule[-0.175pt]{0.723pt}{0.350pt}}
\put(852,429){\rule[-0.175pt]{0.723pt}{0.350pt}}
\put(855,430){\rule[-0.175pt]{0.723pt}{0.350pt}}
\put(858,431){\rule[-0.175pt]{0.723pt}{0.350pt}}
\put(861,432){\rule[-0.175pt]{0.530pt}{0.350pt}}
\put(863,433){\rule[-0.175pt]{0.530pt}{0.350pt}}
\put(865,434){\rule[-0.175pt]{0.530pt}{0.350pt}}
\put(867,435){\rule[-0.175pt]{0.530pt}{0.350pt}}
\put(869,436){\rule[-0.175pt]{0.530pt}{0.350pt}}
\put(872,437){\rule[-0.175pt]{0.578pt}{0.350pt}}
\put(874,438){\rule[-0.175pt]{0.578pt}{0.350pt}}
\put(876,439){\rule[-0.175pt]{0.578pt}{0.350pt}}
\put(879,440){\rule[-0.175pt]{0.578pt}{0.350pt}}
\put(881,441){\rule[-0.175pt]{0.578pt}{0.350pt}}
\put(884,442){\rule[-0.175pt]{0.723pt}{0.350pt}}
\put(887,443){\rule[-0.175pt]{0.723pt}{0.350pt}}
\put(890,444){\rule[-0.175pt]{0.723pt}{0.350pt}}
\put(893,445){\rule[-0.175pt]{0.723pt}{0.350pt}}
\put(896,446){\rule[-0.175pt]{0.530pt}{0.350pt}}
\put(898,447){\rule[-0.175pt]{0.530pt}{0.350pt}}
\put(900,448){\rule[-0.175pt]{0.530pt}{0.350pt}}
\put(902,449){\rule[-0.175pt]{0.530pt}{0.350pt}}
\put(904,450){\rule[-0.175pt]{0.530pt}{0.350pt}}
\put(907,451){\rule[-0.175pt]{0.723pt}{0.350pt}}
\put(910,452){\rule[-0.175pt]{0.723pt}{0.350pt}}
\put(913,453){\rule[-0.175pt]{0.723pt}{0.350pt}}
\put(916,454){\rule[-0.175pt]{0.723pt}{0.350pt}}
\put(919,455){\rule[-0.175pt]{0.723pt}{0.350pt}}
\put(922,456){\rule[-0.175pt]{0.723pt}{0.350pt}}
\put(925,457){\rule[-0.175pt]{0.723pt}{0.350pt}}
\put(928,458){\rule[-0.175pt]{0.723pt}{0.350pt}}
\put(931,459){\rule[-0.175pt]{0.578pt}{0.350pt}}
\put(933,460){\rule[-0.175pt]{0.578pt}{0.350pt}}
\put(935,461){\rule[-0.175pt]{0.578pt}{0.350pt}}
\put(938,462){\rule[-0.175pt]{0.578pt}{0.350pt}}
\put(940,463){\rule[-0.175pt]{0.578pt}{0.350pt}}
\put(943,464){\rule[-0.175pt]{0.662pt}{0.350pt}}
\put(945,465){\rule[-0.175pt]{0.662pt}{0.350pt}}
\put(948,466){\rule[-0.175pt]{0.662pt}{0.350pt}}
\put(951,467){\rule[-0.175pt]{0.662pt}{0.350pt}}
\put(954,468){\rule[-0.175pt]{0.723pt}{0.350pt}}
\put(957,469){\rule[-0.175pt]{0.723pt}{0.350pt}}
\put(960,470){\rule[-0.175pt]{0.723pt}{0.350pt}}
\put(963,471){\rule[-0.175pt]{0.723pt}{0.350pt}}
\put(966,472){\rule[-0.175pt]{0.723pt}{0.350pt}}
\put(969,473){\rule[-0.175pt]{0.723pt}{0.350pt}}
\put(972,474){\rule[-0.175pt]{0.723pt}{0.350pt}}
\put(975,475){\rule[-0.175pt]{0.723pt}{0.350pt}}
\put(978,476){\rule[-0.175pt]{0.662pt}{0.350pt}}
\put(980,477){\rule[-0.175pt]{0.662pt}{0.350pt}}
\put(983,478){\rule[-0.175pt]{0.662pt}{0.350pt}}
\put(986,479){\rule[-0.175pt]{0.662pt}{0.350pt}}
\put(989,480){\rule[-0.175pt]{0.723pt}{0.350pt}}
\put(992,481){\rule[-0.175pt]{0.723pt}{0.350pt}}
\put(995,482){\rule[-0.175pt]{0.723pt}{0.350pt}}
\put(998,483){\rule[-0.175pt]{0.723pt}{0.350pt}}
\put(1001,484){\rule[-0.175pt]{0.964pt}{0.350pt}}
\put(1005,485){\rule[-0.175pt]{0.964pt}{0.350pt}}
\put(1009,486){\rule[-0.175pt]{0.964pt}{0.350pt}}
\put(1013,487){\rule[-0.175pt]{0.723pt}{0.350pt}}
\put(1016,488){\rule[-0.175pt]{0.723pt}{0.350pt}}
\put(1019,489){\rule[-0.175pt]{0.723pt}{0.350pt}}
\put(1022,490){\rule[-0.175pt]{0.723pt}{0.350pt}}
\put(1025,491){\rule[-0.175pt]{0.883pt}{0.350pt}}
\put(1028,492){\rule[-0.175pt]{0.883pt}{0.350pt}}
\put(1032,493){\rule[-0.175pt]{0.883pt}{0.350pt}}
\put(1035,494){\rule[-0.175pt]{0.723pt}{0.350pt}}
\put(1039,495){\rule[-0.175pt]{0.723pt}{0.350pt}}
\put(1042,496){\rule[-0.175pt]{0.723pt}{0.350pt}}
\put(1045,497){\rule[-0.175pt]{0.723pt}{0.350pt}}
\put(1048,498){\rule[-0.175pt]{0.964pt}{0.350pt}}
\put(1052,499){\rule[-0.175pt]{0.964pt}{0.350pt}}
\put(1056,500){\rule[-0.175pt]{0.964pt}{0.350pt}}
\put(1060,501){\rule[-0.175pt]{0.964pt}{0.350pt}}
\put(1064,502){\rule[-0.175pt]{0.964pt}{0.350pt}}
\put(1068,503){\rule[-0.175pt]{0.964pt}{0.350pt}}
\put(1072,504){\rule[-0.175pt]{0.883pt}{0.350pt}}
\put(1075,505){\rule[-0.175pt]{0.883pt}{0.350pt}}
\put(1079,506){\rule[-0.175pt]{0.883pt}{0.350pt}}
\put(1082,507){\rule[-0.175pt]{1.445pt}{0.350pt}}
\put(1089,508){\rule[-0.175pt]{1.445pt}{0.350pt}}
\put(1095,509){\rule[-0.175pt]{0.964pt}{0.350pt}}
\put(1099,510){\rule[-0.175pt]{0.964pt}{0.350pt}}
\put(1103,511){\rule[-0.175pt]{0.964pt}{0.350pt}}
\put(1107,512){\rule[-0.175pt]{1.325pt}{0.350pt}}
\put(1112,513){\rule[-0.175pt]{1.325pt}{0.350pt}}
\put(1118,514){\rule[-0.175pt]{1.445pt}{0.350pt}}
\put(1124,515){\rule[-0.175pt]{1.445pt}{0.350pt}}
\put(1130,516){\rule[-0.175pt]{1.445pt}{0.350pt}}
\put(1136,517){\rule[-0.175pt]{1.445pt}{0.350pt}}
\put(1142,518){\rule[-0.175pt]{1.445pt}{0.350pt}}
\put(1148,519){\rule[-0.175pt]{1.445pt}{0.350pt}}
\put(1154,520){\rule[-0.175pt]{2.650pt}{0.350pt}}
\put(1165,521){\rule[-0.175pt]{2.891pt}{0.350pt}}
\put(1177,522){\rule[-0.175pt]{11.322pt}{0.350pt}}
\put(1224,521){\rule[-0.175pt]{2.891pt}{0.350pt}}
\put(1236,520){\rule[-0.175pt]{1.325pt}{0.350pt}}
\put(1241,519){\rule[-0.175pt]{1.325pt}{0.350pt}}
\put(1247,518){\rule[-0.175pt]{0.964pt}{0.350pt}}
\put(1251,517){\rule[-0.175pt]{0.964pt}{0.350pt}}
\put(1255,516){\rule[-0.175pt]{0.964pt}{0.350pt}}
\put(1259,515){\rule[-0.175pt]{0.723pt}{0.350pt}}
\put(1262,514){\rule[-0.175pt]{0.723pt}{0.350pt}}
\put(1265,513){\rule[-0.175pt]{0.723pt}{0.350pt}}
\put(1268,512){\rule[-0.175pt]{0.723pt}{0.350pt}}
\put(1271,511){\rule[-0.175pt]{0.662pt}{0.350pt}}
\put(1273,510){\rule[-0.175pt]{0.662pt}{0.350pt}}
\put(1276,509){\rule[-0.175pt]{0.662pt}{0.350pt}}
\put(1279,508){\rule[-0.175pt]{0.662pt}{0.350pt}}
\put(1282,507){\rule[-0.175pt]{0.482pt}{0.350pt}}
\put(1284,506){\rule[-0.175pt]{0.482pt}{0.350pt}}
\put(1286,505){\rule[-0.175pt]{0.482pt}{0.350pt}}
\put(1288,504){\rule[-0.175pt]{0.482pt}{0.350pt}}
\put(1290,503){\rule[-0.175pt]{0.482pt}{0.350pt}}
\put(1292,502){\rule[-0.175pt]{0.482pt}{0.350pt}}
\put(1294,501){\rule[-0.175pt]{0.413pt}{0.350pt}}
\put(1295,500){\rule[-0.175pt]{0.413pt}{0.350pt}}
\put(1297,499){\rule[-0.175pt]{0.413pt}{0.350pt}}
\put(1299,498){\rule[-0.175pt]{0.413pt}{0.350pt}}
\put(1300,497){\rule[-0.175pt]{0.413pt}{0.350pt}}
\put(1302,496){\rule[-0.175pt]{0.413pt}{0.350pt}}
\put(1304,495){\rule[-0.175pt]{0.413pt}{0.350pt}}
\put(1305,494){\usebox{\plotpoint}}
\put(1307,493){\usebox{\plotpoint}}
\put(1308,492){\usebox{\plotpoint}}
\put(1310,491){\usebox{\plotpoint}}
\put(1311,490){\usebox{\plotpoint}}
\put(1312,489){\usebox{\plotpoint}}
\put(1314,488){\usebox{\plotpoint}}
\put(1315,487){\usebox{\plotpoint}}
\put(1316,486){\usebox{\plotpoint}}
\put(1318,485){\usebox{\plotpoint}}
\put(1319,484){\usebox{\plotpoint}}
\put(1320,483){\usebox{\plotpoint}}
\put(1321,482){\usebox{\plotpoint}}
\put(1322,481){\usebox{\plotpoint}}
\put(1323,480){\usebox{\plotpoint}}
\put(1324,479){\usebox{\plotpoint}}
\put(1325,478){\usebox{\plotpoint}}
\put(1326,477){\usebox{\plotpoint}}
\put(1327,476){\usebox{\plotpoint}}
\put(1328,475){\usebox{\plotpoint}}
\put(1329,472){\usebox{\plotpoint}}
\put(1330,471){\usebox{\plotpoint}}
\put(1331,470){\usebox{\plotpoint}}
\put(1332,469){\usebox{\plotpoint}}
\put(1333,468){\usebox{\plotpoint}}
\put(1334,467){\usebox{\plotpoint}}
\put(1335,466){\usebox{\plotpoint}}
\put(1336,465){\usebox{\plotpoint}}
\put(1337,464){\usebox{\plotpoint}}
\put(1338,463){\usebox{\plotpoint}}
\put(1339,462){\usebox{\plotpoint}}
\put(1340,461){\usebox{\plotpoint}}
\put(1341,459){\usebox{\plotpoint}}
\put(1342,458){\usebox{\plotpoint}}
\put(1343,456){\usebox{\plotpoint}}
\put(1344,455){\usebox{\plotpoint}}
\put(1345,454){\usebox{\plotpoint}}
\put(1346,452){\usebox{\plotpoint}}
\put(1347,451){\usebox{\plotpoint}}
\put(1348,450){\usebox{\plotpoint}}
\put(1349,448){\usebox{\plotpoint}}
\put(1350,447){\usebox{\plotpoint}}
\put(1351,446){\usebox{\plotpoint}}
\put(1352,445){\usebox{\plotpoint}}
\put(1353,443){\rule[-0.175pt]{0.350pt}{0.401pt}}
\put(1354,441){\rule[-0.175pt]{0.350pt}{0.401pt}}
\put(1355,440){\rule[-0.175pt]{0.350pt}{0.401pt}}
\put(1356,438){\rule[-0.175pt]{0.350pt}{0.401pt}}
\put(1357,436){\rule[-0.175pt]{0.350pt}{0.401pt}}
\put(1358,435){\rule[-0.175pt]{0.350pt}{0.401pt}}
\put(1359,433){\rule[-0.175pt]{0.350pt}{0.401pt}}
\put(1360,431){\rule[-0.175pt]{0.350pt}{0.401pt}}
\put(1361,430){\rule[-0.175pt]{0.350pt}{0.401pt}}
\put(1362,428){\rule[-0.175pt]{0.350pt}{0.401pt}}
\put(1363,426){\rule[-0.175pt]{0.350pt}{0.401pt}}
\put(1364,425){\rule[-0.175pt]{0.350pt}{0.401pt}}
\put(1365,422){\rule[-0.175pt]{0.350pt}{0.548pt}}
\put(1366,420){\rule[-0.175pt]{0.350pt}{0.548pt}}
\put(1367,418){\rule[-0.175pt]{0.350pt}{0.548pt}}
\put(1368,415){\rule[-0.175pt]{0.350pt}{0.548pt}}
\put(1369,413){\rule[-0.175pt]{0.350pt}{0.548pt}}
\put(1370,411){\rule[-0.175pt]{0.350pt}{0.548pt}}
\put(1371,409){\rule[-0.175pt]{0.350pt}{0.548pt}}
\put(1372,406){\rule[-0.175pt]{0.350pt}{0.548pt}}
\put(1373,404){\rule[-0.175pt]{0.350pt}{0.548pt}}
\put(1374,402){\rule[-0.175pt]{0.350pt}{0.548pt}}
\put(1375,400){\rule[-0.175pt]{0.350pt}{0.547pt}}
\put(1376,397){\rule[-0.175pt]{0.350pt}{0.622pt}}
\put(1377,394){\rule[-0.175pt]{0.350pt}{0.622pt}}
\put(1378,392){\rule[-0.175pt]{0.350pt}{0.622pt}}
\put(1379,389){\rule[-0.175pt]{0.350pt}{0.622pt}}
\put(1380,387){\rule[-0.175pt]{0.350pt}{0.622pt}}
\put(1381,384){\rule[-0.175pt]{0.350pt}{0.622pt}}
\put(1382,381){\rule[-0.175pt]{0.350pt}{0.622pt}}
\put(1383,379){\rule[-0.175pt]{0.350pt}{0.622pt}}
\put(1384,376){\rule[-0.175pt]{0.350pt}{0.622pt}}
\put(1385,374){\rule[-0.175pt]{0.350pt}{0.622pt}}
\put(1386,371){\rule[-0.175pt]{0.350pt}{0.622pt}}
\put(1387,369){\rule[-0.175pt]{0.350pt}{0.622pt}}
\put(1388,365){\rule[-0.175pt]{0.350pt}{0.823pt}}
\put(1389,362){\rule[-0.175pt]{0.350pt}{0.823pt}}
\put(1390,358){\rule[-0.175pt]{0.350pt}{0.823pt}}
\put(1391,355){\rule[-0.175pt]{0.350pt}{0.823pt}}
\put(1392,351){\rule[-0.175pt]{0.350pt}{0.823pt}}
\put(1393,348){\rule[-0.175pt]{0.350pt}{0.823pt}}
\put(1394,345){\rule[-0.175pt]{0.350pt}{0.823pt}}
\put(1395,341){\rule[-0.175pt]{0.350pt}{0.823pt}}
\put(1396,338){\rule[-0.175pt]{0.350pt}{0.823pt}}
\put(1397,334){\rule[-0.175pt]{0.350pt}{0.823pt}}
\put(1398,331){\rule[-0.175pt]{0.350pt}{0.823pt}}
\put(1399,328){\rule[-0.175pt]{0.350pt}{0.823pt}}
\put(1400,323){\rule[-0.175pt]{0.350pt}{1.161pt}}
\put(1401,318){\rule[-0.175pt]{0.350pt}{1.161pt}}
\put(1402,313){\rule[-0.175pt]{0.350pt}{1.161pt}}
\put(1403,308){\rule[-0.175pt]{0.350pt}{1.161pt}}
\put(1404,303){\rule[-0.175pt]{0.350pt}{1.161pt}}
\put(1405,299){\rule[-0.175pt]{0.350pt}{1.161pt}}
\put(1406,294){\rule[-0.175pt]{0.350pt}{1.161pt}}
\put(1407,289){\rule[-0.175pt]{0.350pt}{1.161pt}}
\put(1408,284){\rule[-0.175pt]{0.350pt}{1.161pt}}
\put(1409,279){\rule[-0.175pt]{0.350pt}{1.161pt}}
\put(1410,275){\rule[-0.175pt]{0.350pt}{1.161pt}}
\put(1411,269){\rule[-0.175pt]{0.350pt}{1.365pt}}
\put(1412,263){\rule[-0.175pt]{0.350pt}{1.365pt}}
\put(1413,258){\rule[-0.175pt]{0.350pt}{1.365pt}}
\put(1414,252){\rule[-0.175pt]{0.350pt}{1.365pt}}
\put(1415,246){\rule[-0.175pt]{0.350pt}{1.365pt}}
\put(1416,241){\rule[-0.175pt]{0.350pt}{1.365pt}}
\put(1417,235){\rule[-0.175pt]{0.350pt}{1.365pt}}
\put(1418,229){\rule[-0.175pt]{0.350pt}{1.365pt}}
\put(1419,224){\rule[-0.175pt]{0.350pt}{1.365pt}}
\put(1420,218){\rule[-0.175pt]{0.350pt}{1.365pt}}
\put(1421,212){\rule[-0.175pt]{0.350pt}{1.365pt}}
\put(1422,207){\rule[-0.175pt]{0.350pt}{1.365pt}}
\put(1423,202){\rule[-0.175pt]{0.350pt}{0.984pt}}
\put(1424,198){\rule[-0.175pt]{0.350pt}{0.984pt}}
\put(1425,194){\rule[-0.175pt]{0.350pt}{0.984pt}}
\put(1426,190){\rule[-0.175pt]{0.350pt}{0.984pt}}
\put(1427,186){\rule[-0.175pt]{0.350pt}{0.984pt}}
\put(1428,182){\rule[-0.175pt]{0.350pt}{0.984pt}}
\put(1429,178){\rule[-0.175pt]{0.350pt}{0.984pt}}
\put(1430,174){\rule[-0.175pt]{0.350pt}{0.984pt}}
\put(1431,170){\rule[-0.175pt]{0.350pt}{0.984pt}}
\put(1432,166){\rule[-0.175pt]{0.350pt}{0.984pt}}
\put(1433,162){\rule[-0.175pt]{0.350pt}{0.984pt}}
\put(1434,158){\rule[-0.175pt]{0.350pt}{0.984pt}}
\put(1435,158){\usebox{\plotpoint}}
\sbox{\plotpoint}{\rule[-0.250pt]{0.500pt}{0.500pt}}%
\put(264,158){\usebox{\plotpoint}}
\put(264,158){\usebox{\plotpoint}}
\put(282,167){\usebox{\plotpoint}}
\put(300,176){\usebox{\plotpoint}}
\put(319,186){\usebox{\plotpoint}}
\put(337,195){\usebox{\plotpoint}}
\put(356,205){\usebox{\plotpoint}}
\put(374,214){\usebox{\plotpoint}}
\put(393,224){\usebox{\plotpoint}}
\put(411,233){\usebox{\plotpoint}}
\put(430,243){\usebox{\plotpoint}}
\put(448,253){\usebox{\plotpoint}}
\put(466,262){\usebox{\plotpoint}}
\put(485,272){\usebox{\plotpoint}}
\put(503,281){\usebox{\plotpoint}}
\put(522,291){\usebox{\plotpoint}}
\put(540,300){\usebox{\plotpoint}}
\put(559,310){\usebox{\plotpoint}}
\put(577,319){\usebox{\plotpoint}}
\put(596,329){\usebox{\plotpoint}}
\put(614,338){\usebox{\plotpoint}}
\put(633,347){\usebox{\plotpoint}}
\put(651,357){\usebox{\plotpoint}}
\put(669,366){\usebox{\plotpoint}}
\put(688,376){\usebox{\plotpoint}}
\put(707,385){\usebox{\plotpoint}}
\put(725,395){\usebox{\plotpoint}}
\put(743,404){\usebox{\plotpoint}}
\put(761,414){\usebox{\plotpoint}}
\put(780,424){\usebox{\plotpoint}}
\put(798,433){\usebox{\plotpoint}}
\put(817,443){\usebox{\plotpoint}}
\put(835,452){\usebox{\plotpoint}}
\put(854,462){\usebox{\plotpoint}}
\put(872,471){\usebox{\plotpoint}}
\put(891,481){\usebox{\plotpoint}}
\put(909,490){\usebox{\plotpoint}}
\put(928,499){\usebox{\plotpoint}}
\put(946,509){\usebox{\plotpoint}}
\put(965,519){\usebox{\plotpoint}}
\put(983,528){\usebox{\plotpoint}}
\put(1002,537){\usebox{\plotpoint}}
\put(1020,547){\usebox{\plotpoint}}
\put(1039,556){\usebox{\plotpoint}}
\put(1057,566){\usebox{\plotpoint}}
\put(1075,576){\usebox{\plotpoint}}
\put(1094,586){\usebox{\plotpoint}}
\put(1112,595){\usebox{\plotpoint}}
\put(1131,604){\usebox{\plotpoint}}
\put(1149,614){\usebox{\plotpoint}}
\put(1167,623){\usebox{\plotpoint}}
\put(1186,633){\usebox{\plotpoint}}
\put(1205,642){\usebox{\plotpoint}}
\put(1223,652){\usebox{\plotpoint}}
\put(1241,661){\usebox{\plotpoint}}
\put(1260,671){\usebox{\plotpoint}}
\put(1278,680){\usebox{\plotpoint}}
\put(1297,689){\usebox{\plotpoint}}
\put(1315,699){\usebox{\plotpoint}}
\put(1334,708){\usebox{\plotpoint}}
\put(1352,718){\usebox{\plotpoint}}
\put(1371,727){\usebox{\plotpoint}}
\put(1389,737){\usebox{\plotpoint}}
\put(1407,747){\usebox{\plotpoint}}
\put(600,-400){Fig. 2.}
\end{picture}

\end{document}